\documentclass[aps,onecolumn,groupedaddress,amsmath,amssymb,superscriptaddress]{revtex4}

\usepackage{graphics}
\usepackage{graphicx}
\usepackage{float}
\usepackage{amsmath}
\usepackage{amsfonts}
\usepackage{amssymb}
\usepackage{pstricks}
\usepackage{relsize}
\usepackage[normalem]{ulem}
\usepackage[caption=false]{subfig}
\usepackage{soul,xcolor}

  \usepackage{color}
  \usepackage{nicefrac}

  \newcommand{\p}{\partial}

\begin{document}

\title{Coexistence of stable branched patterns  in anisotropic
inhomogeneous systems}

\author{B. Kaoui$^1$, A. Guckenberger$^1$,  A. Krekhov$^{1,2}$, F. Ziebert$^{1,3}$ and W. Zimmermann$^{1}$}

\address{$^1$Theoretische Physik, Universit\"at Bayreuth, 95440 Bayreuth, Germany \\
$^2$Max-Planck-Institute for Dynamics and Self-Organization, 37077 G\"ottingen, Germany \\
$^3$Physikalisches Institut, Albert-Ludwigs-Universit\"at Freiburg, 79104 Freiburg, Germany
}
\email{walter.zimmermann@uni-bayreuth.de
}

\date{\today}

\begin{abstract}
A new class of pattern forming systems
is identified and investigated: anisotropic systems 
that are spatially  inhomogeneous along the direction perpendicular to the preferred one. 
By studying the generic amplitude equation of this new class and a model equation, we show that
branched stripe patterns emerge, which for a given parameter set are stable
within a band of different wavenumbers and different numbers of branching points (defects).
Moreover, the branched patterns and unbranched ones (defect-free stripes) coexist
over a finite parameter range. We propose two systems where 
this generic scenario 
can be found experimentally, surface wrinkling on elastic substrates and 
electroconvection in nematic liquid crystals, and relate them to the findings from
the amplitude equation. 
\end{abstract}

%\pacs{????} 

\maketitle

\section{Introduction}\label{sec: intro}
 
Pattern formation is one of the most fascinating and intriguing phenomena in nature
\cite{Ball:98,CrossHo}. It takes place in a wide variety of physical, chemical and
biological systems and on disparate spatial and temporal scales, for example,
convection phenomena in geoscience \cite{CrossHo,Lappa:2010} 
or in liquid crystals \cite{Kramer:96,Zimmermann:88.3,Kramer:95.1}, 
environmental patterns \cite{Feingold:2010.1,Goehring:2009.1,Meron:2012.2}, 
or patterns in chemical reactions \cite{Kapral:1995,Mikhailov:2006.1} and
bacterial colonies \cite{BenJacob:1994.1}. 
In some circumstances pattern formation is undesired,
for instance, the formation of spiral waves leading to cardiac arrhytmias in the heart muscle \cite{Jalife:2009}.
In other contexts pattern formation is even essential for the
functioning  of a system, e.g.~in embryo development \cite{Wieschaus:2005.1} 
or when designing surface wrinkling patterns 
to fabricate nanometer-scale structures \cite{bowden,Groenewold_Rev}. 
The mechanisms leading to the same type of pattern 
in different systems are obviously very diverse. 
Nevertheless, patterns 
occuring in systems  of the same symmetry share common qualitative properties
that can be described by universal amplitude equations for the envelope of
periodic patterns
\cite{Newell:1992.1,Zimmermann:88.3,Kramer:95.1,CrossHo}.

Generating, modifying or eliminating patterns 
hence either requires a profound understanding of
 the pattern formation mechanism in each specific system, or
complementary, of the  universal properties of a class of patterns and their 
response to symmetry breakings.
Common pattern interventions are
feed back control \cite{Mikhailov:2006.1} and 
symmetry breaking via spatial, temporal or spatio-temporal modulations
\cite{Lowe:83.1,Coullet:86.2,Zimmermann:93.3,PeterR:2005.1,Freund:2011.1,Meron:2012.1,Bodenschatz:2012.1,
Rehberg:88.1,Riecke:88.1,Walgraef:88.1,
Kramer:2004.1,Schuler:2004.1}. 
To mention two interesting scenarios, 
spatial forcing near resonance 
(between the forcing and the natural wavelength) 
can lead to so-called incommensurate patterns 
\cite{Lowe:83.1,Coullet:86.2}, while
symmetry breaking via long-wave spatial modulations can 
render stationary patterns time dependent
\cite{Rehberg:91.2,Zimmermann:96.3,Zimmermann:2002.2,Zimmermann:2002.3}.
The response of patterns in quasi one-dimensional (1D)
systems is now well established, however truely 
2D scenarios, like the interplay of an anisotropy and a modulation in different directions, 
are yet fully unexplored.

\begin{figure}[htb]
\begin{center}
\includegraphics[width=0.7\textwidth]{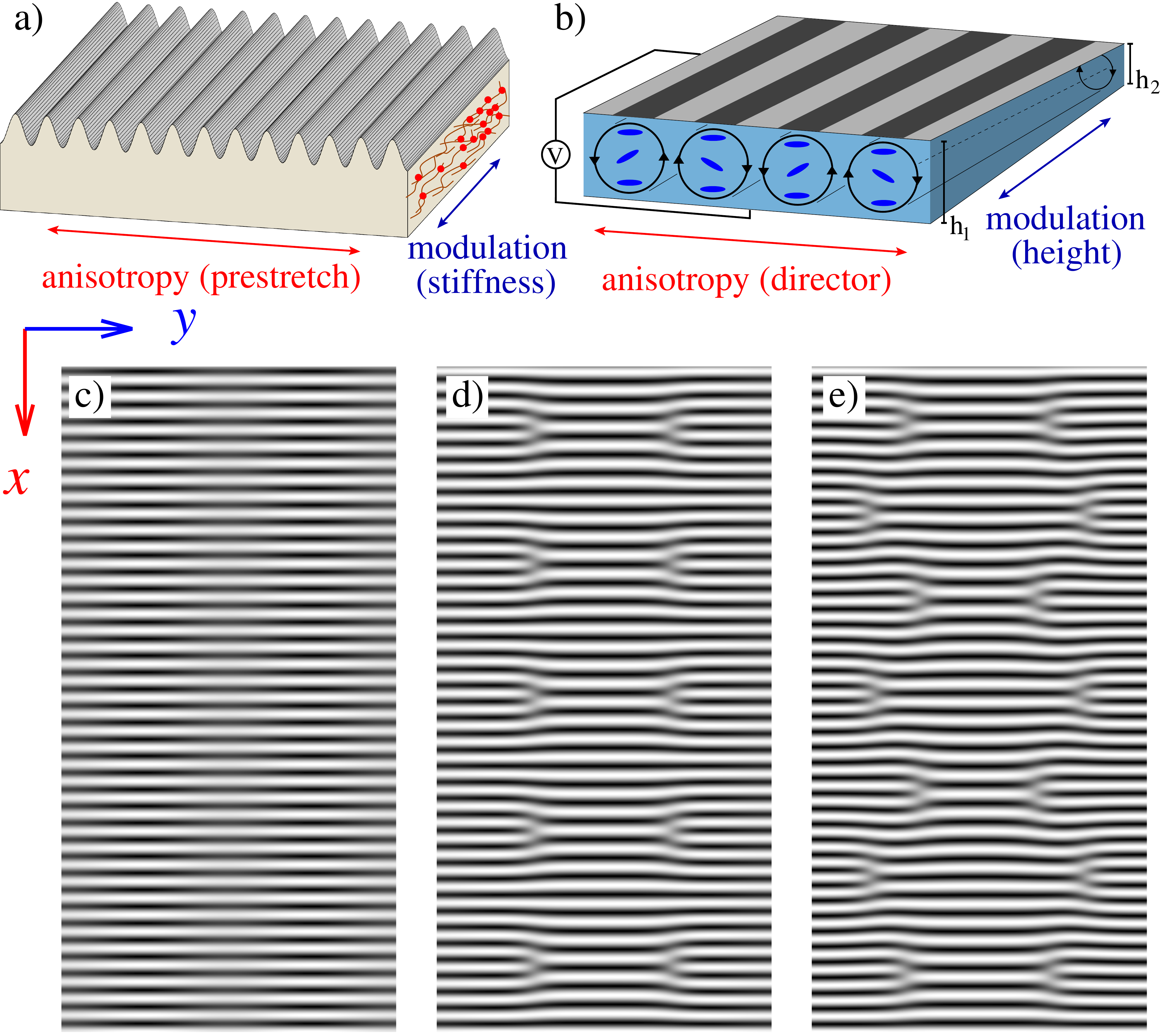}
\end{center}
\caption{\label{fig:Figure01}
%(Color online)
 In (a)  a wrinkle forming elastic system is sketched with a spatially varying 
 stiffness and in (b) electroconvection in nematic liquid crystals with a spatially varying height
 of the convection cell.
 Parts (c)-(e) show selected characteristic patterns 
for the proposed system class. They are stable at identical parameters and
have been obtained numerically via equation~(\ref{Ampli_A})
for the parameter set $\hat{q}_0=1$, $k_m=0.05$, $M=0.1$, $\varepsilon=0.08$
and for different initial conditions by using the pseudo-spectral method \cite{Kopriva}.
Each picture is a cutout of a larger domain 
($L_x=80 \times 2\pi/\hat q_0$ and $L_y = 2\pi/k_m$). 
}
\end{figure}

In this paper we identify and analyze a new 
class of quasi-2D {\it anisotropic inhomogeneous} pattern forming systems.
 The wave vector of the patterns 
lies close to  $\hat {\bf q}_0=(\hat q_0,0)$ along the preferred $x$-direction (anisotropy), 
as for the two experimental systems sketched in figure~\ref{fig:Figure01}:  
(a) a spatially modulated version of the wrinkle 
forming system \cite{Groenewold_Rev,Glatz2015} and 
(b) a modulated version of electroconvection (EHC) in nematic liquid crystals \cite{Kramer:95.1}.
 By a modulation we break in such systems the  translational symmetry
along the perpendicular $y$-direction, causing a variation
of the pattern's natural wavenumber $\hat q_0$. 
This can be accomplished by varying the elasticity in the wrinkle forming system
or the  height  of the electroconvection cell, respectively. 
In this  class of anisotropic systems we find
straight stripes, cf.~figure~\ref{fig:Figure01}(c), 
that are stable for wavenumbers in a finite range around 
$\hat {q}_0$, similar as in homogeneous anisotropic systems. 
In addition, however, we find a whole family of stable branched patterns 
as shown in figure~\ref{fig:Figure01}(d) and (e).
Surprisingly, they have -- at identical parameters -- 
different characteristic wave numbers and include different numbers of branching points. 
Moreover, the branched patterns coexist with the straight stripes in a wide parameter range.
This behavior is a non-trivial generalization of the wave number bands (Eckhaus bands)
for homogeneous systems 
\cite{Eckhaus:65,Zimmermann:85.1,Lowe:85.2,Zimmermann:85.2,Dominguez-Lerma:86.1,Riecke:86.1,Kramer:1986.1}
-- a well established concept 
and experimentally verified e.g.~in EHC \cite{Lowe:85.2} and axisymmetric 
Taylor vortex flow \cite{Dominguez-Lerma:86.1,Riecke:86.1} --
to inhomogeneous systems and multiple patterns.
In the following we present the universal amplitude equation of this new symmetry class
of patterns and analyze its solutions.

\section{Model and generic amplitude equation}
A generic model for the formation of stationary periodic patterns in anisotropic 
-- but homogeneous --  2D systems, described by a field $u(x,y,t)$, has been proposed in Ref.~\cite{Kramer:1986.1}.
One interpretation of the field $u(x,y,t)$ is, that it  describes the (small) lateral displacement of a thin
elastic plate extended in the $x$-$y$ plane, loaded along the $x$ and $y$ direction and supported by
an elastic medium, similar as in wrinkling systems  (whereby the in-plane elastic deformations are neglected) \cite{Kramer:1986.1}. 
Here we generalize this model to an inhomogeneous situation 
by modulating the pattern's preferred natural wavenumber $\hat {q}_0$ 
along the direction perpendicular to the anisotropy (the $x$-direction),
 \begin{align}
\label{q0mod}
 q_0(y)= \hat q_{0} + M \cos(k_my)\,,
\end{align}
with an amplitude $M$ and a  wavenumber $k_m$ considerably smaller than $\hat q_0$. Then the dynamics 
of the patterns, described by 
the real field $u(x,y,t)$
is goverend by
\begin{align}
\label{SHanisomod}
 \partial_t u&= \left[ \varepsilon - \left(q_0^2(y)+\Delta\right)^2\right] u-W\partial_x^2 \partial_y^2 u
-c\partial_y^4 u-u^3 \nonumber\\
&\qquad \qquad -2 (\partial_y u) \partial_y \left[q_0^2(y)] - u \partial_y^2[q_0^2(y)\right]\,.
\end{align}
The first line corresponds to the original model in Ref.~\cite{Kramer:1986.1}
and 
the second line includes  contributions due to the modulation $q_0(y)$. 
Equation~(\ref{SHanisomod}) is  a representative of the here-identified 
symmetry class. It can be directly linked to the elastic wrinkle-forming system via an appropriate rescaling \cite{Kramer:1986.1}: 
$q_0^4=\frac{\kappa}{\lambda_1}$ relates the critical wave number
to the bending stiffness $\lambda_1$ of the hard layer and the
elastic modulus $\kappa$ of the substrate. 
The control parameter $\varepsilon = \left(1 - \frac{\mu_{1,c}^2}{\mu^2 _ 1}\right)q_0^4$
is related to the critical compression $\mu_{1,c}=2\sqrt{\kappa\lambda_1}$. 
Also note that, similar to the homogeneous version \cite{Kramer:1986.1} and the wrinkle system,
equation~(\ref{SHanisomod}) can be derived from a functional as described in the appendix.
For the following, we have chosen
$W=1$, $c = 0.5$ and $\hat{q}_0=1$ that favor straight stripes formation parallel to the $y$-direction for the homogeneous case. 

Close to the onset of  supercritically bifurcating  patterns 
their generic (system-independent) properties may be described by a nonlinear dynamical equation for 
the complex envelope $A(x,y,t)$, considering 
$u(x,y,t) =   A e^{i\hat q_0 x} +  A^\ast  e^{-i\hat q_0 x}$
with the scaling $u \propto \varepsilon^{1/2}$ (the star denotes the complex conjugate).
This reduction method,  the so-called multiple scale analysis, is
well established for supercritical bifurcations in 2D homogeneous isotropic
\cite{Newell:1969.1,CrossHo} and anisotropic systems \cite{Kramer:1986.1,Zimmermann:88.3}
or 1D inhomogeneous ones \cite{Zimmermann:96.3}.
Following very closely the appendix in Ref.~\cite{CrossHo} with the same intermediate
scaling for time and space one obtains
with   $M \propto  \varepsilon^{1/2}$ 
and 
$k_m \propto \varepsilon^{1/2}$ via the 
multiple scale analysis from the system described by equation~(\ref{SHanisomod}) the  generic
amplitude equation for the  here-identified universality class of patterns in anisotropic systems
with a spatially varying natural wave number $q_0(y)$:
\begin{align}
 \label{Ampli_A}
 \partial_t A &= \left[ \varepsilon  +\xi_x^2 \partial_x^2  +
\xi_y^2 \partial_y^2  \right] A - g \lvert A\rvert ^2 A\, 
\nonumber \\ & \qquad
 -i 8 \hat q_0^2 M \cos(k_m y) \partial_x A - 4 \hat q_0^2 M^2 \cos^2(k_m y) A \,.
\end{align}
Upon derivation from equation~(\ref{SHanisomod}) one obtains the relations
$\xi_x^2=4\hat q_0^2$, $\xi_y^2=W \hat q_0^2$ and $g=3$.
The generic phenomenon of stripes' branching, 
as detailed below, is induced in equation~(\ref{Ampli_A}) by the term $\propto M\partial_x A$.
The term $\propto M^2A$ mainly leads to quantitative modifications.
 It should be noted that the first line of equation~(\ref{Ampli_A}) is exactly the
equation for the amplitude of stripe patterns in unmodulated EHC \cite{Kramer:1986.1,Zimmermann:88.3}, 
the second anisotropic system suggested for experimental investigations
of phenomena described here.
For all specific pattern forming systems in the considered universality class, 
equation~(\ref{Ampli_A}) can be derived via perturbation techniques 
from the basic equations (as for example from equation (\ref{SHanisomod}))
and it has always the {\it same} form ---  
only the coefficients will depend on the specific system. 
The presented  results are obtained via the universal equation (\ref{Ampli_A}).

\section{Results and discussion} 
The modulation of the natural wavenumber, see equation~(\ref{q0mod}),
alters the bifurcation from the  basic state 
($u=A=0$)
towards stationary periodic solutions of %of Eq.~(\ref{Ampli_A}), 
the form  $A(x,y)=e^{i Q x}H(y)$.  
The stability of the latter  and of $A=0$ with respect to small perturbations
$v(x,y,t)=v_1(y)e^{\sigma t + iKx}+v_2^*(y)e^{\sigma^* t - iKx}$
can be determined via the ansatz
$A(x, y,t) = e^{i Q x} \left[ H(y) + v(x,y,t)\right]$,
followed by a linearization of equation~(\ref{Ampli_A}) 
with respect to small $v$. An expansion of $H(y)$ and $v_{1,2}(y)$ in Fourier modes 
results in an eigenvalue  problem for  $\sigma$, whereby solutions are stable when
all eigenvalues have negative real parts.
Note that in $v(x,y,t)$ we consider only perturbations along the $x$-direction, 
as perturbations parallel to the stripes (along $y$) have been numerically found to have no effect.

\begin{figure}[htb]
\begin{center}
\includegraphics[width=0.33\textwidth,angle=-90]{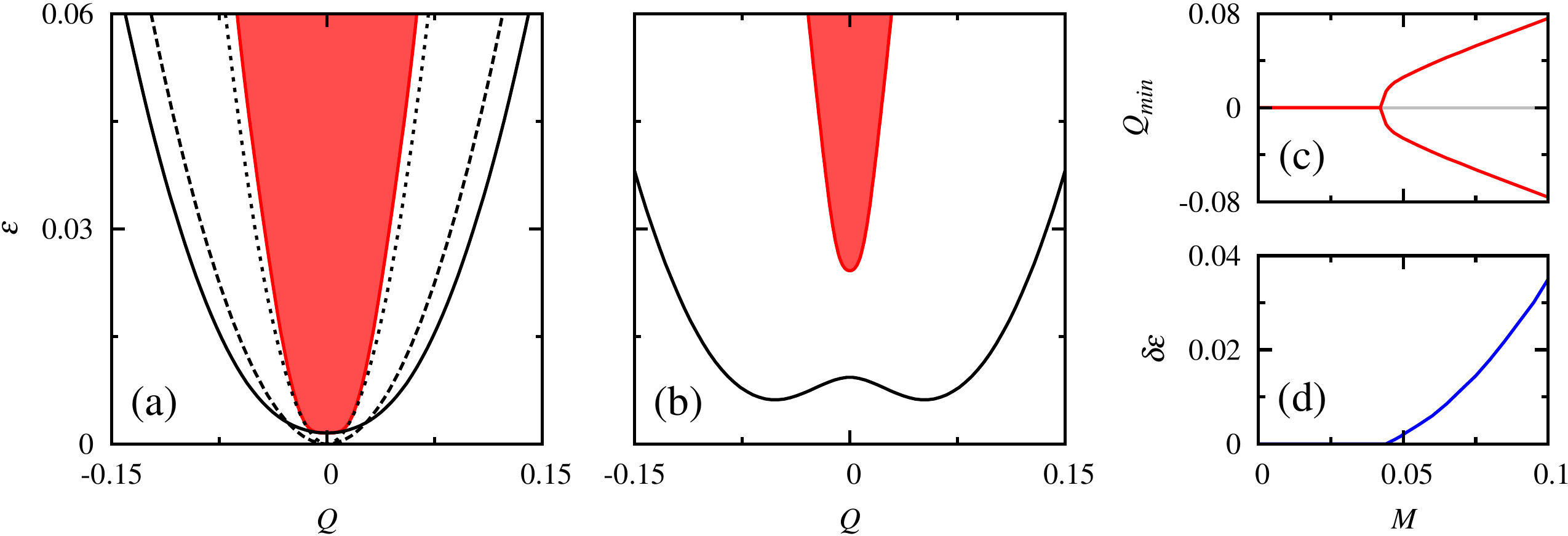}
\end{center}
\vspace{-2mm}
\caption{\label{Ampli_stab} 
%(Color online).  
(a) For small modulation amplitude, $M=0.03$, 
the neutral curve (solid), above which stripes exist, 
and the Eckhaus stability region (in red), 
where stripes are stable,
have the classical single-well shape with a minimum at $Q=0$.
For $M=0$ the neutral (dashed) and the Eckhaus (dotted) curves 
are given for comparison. 
(b) With $M=0.075$ beyond a critical value 
(here $M_c\simeq 0.0459$), 
the neutral curve develops a double-well shape (solid line). 
The Eckhaus stable region for straight stripe patterns (in red) shrinks and detaches 
from the neutral curve: a gap develops, where only branched stripes are stable. 
Within the red region, stable unbranched and branched stripes coexist. 
(c) The minima of the neutral curve, $Q_{min}$,
and (d) the gap $\delta \varepsilon$  between the neutral and the stable range
%are given
as a function of $M$. 
Other parameters: $\hat q_{0}=1$ and $k_m=0.1$.
}
\vspace{-2mm}
\end{figure} 

 For the basic state $A=0$ the stability condition, $\Re \left[\sigma_{max}(\varepsilon,Q)\right]=0$, 
determines the neutral curve $\varepsilon_{\rm N}(Q)$,
above which  stationary periodic solutions, $A=e^{iQx}H(y)$,
of finite amplitude do exist.
The neutral curves  for the unmodulated
case $M=0$ and for a small modulation $M=0.03$ are plotted in figure~\ref{Ampli_stab}(a) with dashed and solid lines, respectively. 
Both curves  have their minimum at $Q=0$ and are symmetric.
However, when increasing $M$, the value of the minimum $\varepsilon_N(0)$ shifts towards 
positive values due to the term $\propto M^2A$ in equation~(\ref{Ampli_A}) that becomes 
larger ($\varepsilon \rightarrow \left[ \varepsilon - 4\hat{q}_0^2M^2\cos^2(k_my)\right]$).
By increasing $M$ further beyond a critical modulation
(here $M_c=0.0459$ for $\hat{q}_0=1$ and $k_m=0.1$)
a very important phenomenon takes place: 
The neutral curve develops two minima at 
finite wave numbers $Q=\pm Q_{min}$, as shown in figure~\ref{Ampli_stab}(b) by the solid line.
The appearence of these two minima  is 
caused by high modulation amplitudes and mainly by the term $\propto M\partial_x A$ in equation~(\ref{Ampli_A}). 
The dependence of $Q_{min} \propto \sqrt{M-M_c}$ on the modulation in the vicinity of $M_c$
resembles a pitchfork bifurcation, cf.~figure~\ref{Ampli_stab}(c). 
 The two new terms in the amplitude equation, caused by the modulation, 
hence widen the neutral curve at small modulations and 
change its shape from a single- to double-well shape via a 
pitchfork-like bifurcation at higher modulations.
The fact that the modulation
along the y-direction causes two minima of the neutral
curve $\varepsilon_N(Q)$, with respect to $Q$ along the $x$ direction, is crucial
for the emergence of  branched patterns.

\begin{figure}[t]
\vspace{-0mm}
\begin{center}
\includegraphics[width=0.22\linewidth,angle=-90]{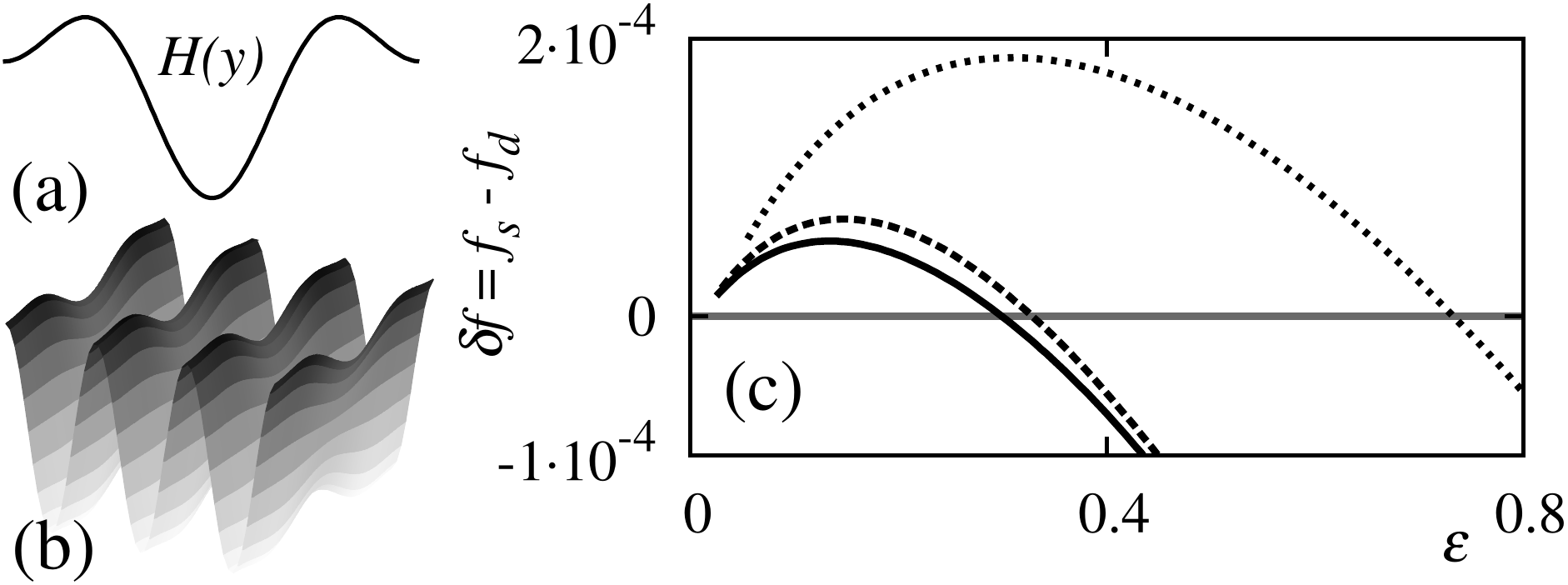}
\end{center}
\vspace{-3mm}
\caption{\label{Potdiff}
(a) The amplitude $H(y)$  of a  modulated stripe pattern given in (b).
(c) The energy density difference between the stripe pattern and one of the possible 
branched patterns, $\delta f=f_s-f_d$, as a function of $\varepsilon$ 
for three different wavenumbers  [$Q=0$ (solid),  $Q=0.0125$ (dashed),  $Q=0.025$ (dotted)]
within the coexistence region.
For small values of $\varepsilon$, branched patterns are energetically preferred, % ($\delta f>0$),
while for large $\varepsilon$ unbranched ones. % ($\delta f>0$). 
The crossover position strongly increases with $Q$. 
Parameters: $\hat{q}_0=1$, $k_m=0.1$, $M=0.075$.
}
\vspace{-5mm}
\end{figure}

Periodic solutions  in the homogeneous case,  $M=0$, have a constant 
amplitude, $H(y)= \sqrt{(\varepsilon -\xi_x^2 Q^2)/g~}$,
and they are linearly stable only above the dotted line in 
figure~\ref{Ampli_stab}(a), the so-called
Eckhaus-stability boundary $\varepsilon_E(Q)$. 
 Note that for 2D anisotropic systems, the $Q$-range of 
stable solutions 
is symmetric with respect to $Q=0$ \cite{Kramer:1986.1},
in contrast to isotropic systems, in which  the zig-zag instability 
occurs for $Q<0$.
For small modulations $M<M_c$, 
$H(y)$ is $2\pi/k_m$-periodic along the $y$-direction
and corresponds to modulated stripes as shown in 
figures~\ref{fig:Figure01}(a) and \ref{Potdiff}(a),(b).
In general, the equation for $H(y)$ may have
 further solutions  as well. 
However, simulations did not show them, so they are either unstable or have higher energies.
The Eckhaus stability range
of the modulated stripes 
is qualitatively unchanged compared to the homogeneous limit, as shown by the red-colored region in
figure~\ref{Ampli_stab}(a): It still touches $\varepsilon_N(Q)$ at $Q=0$. % ($M=0$).
However, quantitatively, the modulation tends 
to narrow the width of the Eckhaus-stability range and thus 
reduces the range of stable stripes.

\begin{figure}[t!]
\begin{center}
\includegraphics[width=0.72\textwidth,angle=0]{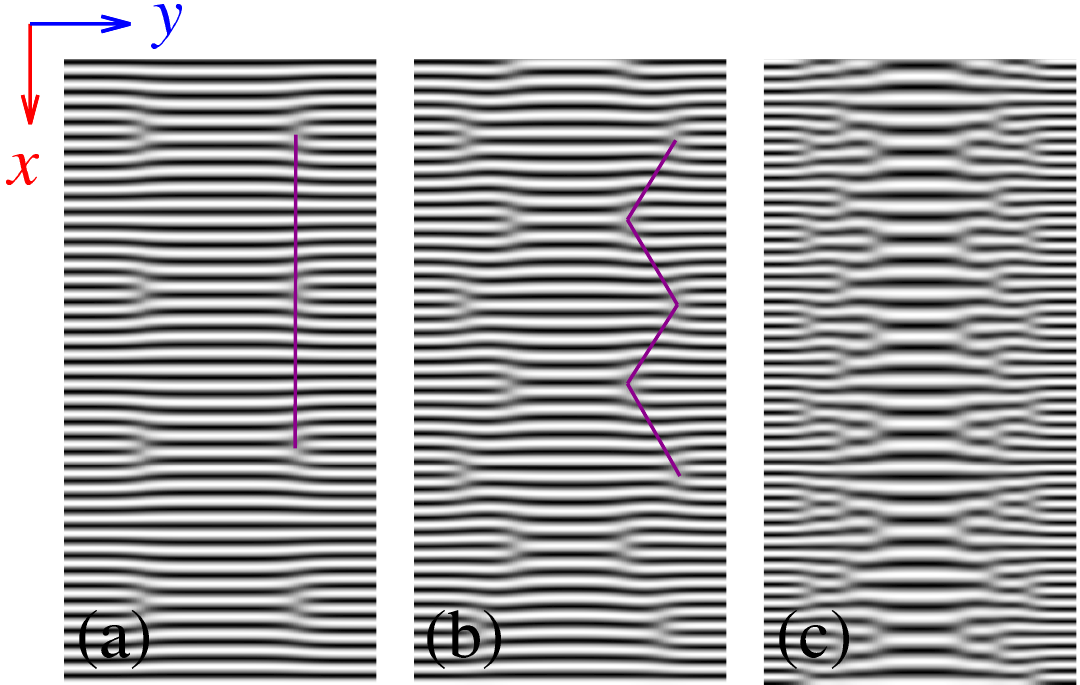}
\end{center}
\vspace{-5mm}
\caption{
%(Color online). 
\label{fig:Figure04}
The branched patterns display different arrangements of defects 
upon increasing the amplitude  $M$: (a) Aligned structure for $M=0.06$,
(b) Zig-zag order for $M=0.1$, 
(c) Cascade-like order for $M=0.3$.
Parameters: $\hat q_{0}=1$, $k_m=0.05$,  $\varepsilon=0.08$, $L_x= 80 \times 2\pi/\hat q_0$,
$L_y=2\pi/k_m$.
}
\end{figure}

Concomitant with 
increasing the modulation beyond the threshold $M_c$
a striking change in the stability scenario occurs:
i) As shown in figure~\ref{Ampli_stab}(b), a gap opens
between the neutral curve (solid line) and the  stability 
range of amplitude modulated straight stripes (red-colored area). 
Withhin  the gap, the stripe solutions $A(x,y)=e^{iQx} H(y)$ are 
unstable and thus cannot be observed anymore. 
The width of this gap  increases nonlinearly with $M - M_c$, 
as shown in figure~\ref{Ampli_stab}(d).
ii) Stable branched stripes, 
like those shown in figures~\ref{fig:Figure01}(d),(e) and figure~\ref{fig:Figure04},
emerge within  a large area above the neutral curve.
iii)  Within the red-colored area in figure~\ref{Ampli_stab}(b), 
the unbranched modulated stripes {\it coexist} 
with branched stripes at identical parameters sets.
To  characterize the coexistence (the first central result) further, we  use the fact that
the dynamical equations~(\ref{SHanisomod}) and (\ref{Ampli_A}) %[and also the analogous equation for $H(y)$]
can be derived from functionals, 
which for stationary patterns take the simple form (appendix):
\begin{align}
\label{funct}
{\cal F} = -\frac{1}{4}\int dxdy ~u^4= -\frac{3}{2} \int dxdy ~ \lvert A\rvert^4 \,.
\end{align}
Using the energy density $f={\cal F}/L_xL_y$ ($L_xL_y$ is the domain area)
one can hence determine
which of the stable patterns is energetically preferred
at a given parameter set (see also p. 868 of Ref. \cite{CrossHo}).
We have calculated the energy densities of
unbranched patterns $f_s$ and of 
branched patterns $f_d$ within the red-colored region in figure~\ref{Ampli_stab}(b). 
Their difference $\delta f=f_s-f_d$ 
is plotted in figure~\ref{Potdiff}(c) as a function of $\varepsilon$ 
and for three different wave numbers $Q$. 
For small  $\varepsilon$, branched patterns have a lower energy and thus are  energetically preferred,
whereas for considerably larger $\varepsilon$, 
unbranched patterns are in turn preferred. 
The crossover can be understood
as
the modulation becomes less important at large 
values of the control parameter
$\varepsilon$.

The second central result
is that for each parameter set out of a large range beyond $\varepsilon_N(Q)$ we find 
always a whole family of stable branched patterns, 
characterized by different wavelengths and distinct numbers
of branching points - and obtained by varying the initial conditions. For example, the two patterns
in parts (d) and (e) in figure~\ref{fig:Figure01} contain five and seven
pairs of branching points at identical parameters. They are both  stable, however,  
the value of the functional of every branched pattern is, in general, different.
This pattern coexistence in inhomogeneous systems is a surprising {\it generalization} 
of the  well established Eckhaus wave number band of 
stable stripe patterns in homogeneous systems
\cite{Eckhaus:65,Zimmermann:85.1,Lowe:85.2,Zimmermann:85.2,Dominguez-Lerma:86.1,Riecke:86.1}. 
There, stable stripes of different wave numbers coexist at an identical parameter set,
as verified in EHC \cite{Lowe:85.2}, buckling \cite{Zimmermann:85.2} and axisymmetric 
Taylor vortex flow \cite{Dominguez-Lerma:86.1,Riecke:86.1}. 
Here, a band of branched patterns with different wavelengths and 
numbers of branching points coexists.
In both cases,
the origin of this coexistence lies in the fact
that transitions between two stable periodic patterns require rather 
high excitations \cite{Zimmermann:85.1} worth to be investigated further.

Another interesting observation 
is due to the branching points (defects) having opposite topological charges 
in each  half period $\pi/k_m$ along the $y$-direction
hence they repel each other.  
As shown in figure~\ref{fig:Figure04}, this repulsion can lead to different 
orderings of the defects
with increasing defect density, 
for instance, the zig-zag ordering shown in figure~\ref{fig:Figure04}(b).
It can even lead to more complex cascade-like ordering, as shown in figure~\ref{fig:Figure04}(c).
However, how the detailed ordering of branching points can be controlled by the magnitude, 
the wavelength and the anharmonicity of the modulation
is an interesting and fundamental question that needs to be adressed in the future.

The phenomena identified here 
are robust and insensitive to the special 
shape of the imposed modulation of the natural pattern wavenumber. 
Studying the generic model, equation~(\ref{SHanisomod}), and imposing 
instead of a harmonic modulation 
either an anharmonic wave-number modulation
or a step-wise one,
we obtain qualitatively the same results \cite{Glatz2015}.
Consequently, instead of harmonic variations, 
which may be challenging to implement in experimental systems,
 a step-like modulations can be used, which is easy to achieve in the wrinkles system by
glueing two elastomer substrates with different elastic 
properties together \cite{SchmidtHW:2012.2,Glatz2015}.

\section{Conclusions}
In conclusion, we identified a new symmetry class of pattern forming systems
that are anisotropic with a perpendicular modulation.
By deriving and analyzing the universal amplitude equation, 
we showed that this class displays interesting scenarios 
-- the emergence of families of stable branched patterns and 
their coexistence with  unbranched patterns --  
that we suggest to verify experimentally 
in two complementary  anisotropic systems:  
First, in  wrinkle forming systems \cite{Groenewold_Rev}, 
where the elasticity of the substrate 
[as sketched in figure~\ref{fig:Figure01}(a)] 
and/or the thickness of the hard layer on top can 
be varied perpendicularly to the pre-stretch (anisotropy) direction,
as recently realized experimentally \cite{SchmidtHW:2012.2,Glatz2015}.
And second, in dissipative electroconvection in nematic liquid crystals 
\cite{Kramer:96,Zimmermann:88.3,Kramer:95.1}, where 
the layer height [as sketched in figure~\ref{fig:Figure01}(b)] 
or the driving frequency are modulated perpendicularly to the nematic alignment. 
For both systems, a thorough theoretical analysis of the
modulated basic equations is feasible, including the derivation of the
universal equation of the patterns envelope, i.e.~equation~(\ref{Ampli_A}). 
There is a vast literature on wrinkles recently (see e. g. Refs.~\cite{bowden,Groenewold_Rev,Glatz2015,Damman:2010.2,He:2012.1,Chen:2010.1}), 
including control strategies for wrinkle formation,
but the question of pattern  {\it coexistence} seems to have not even been raised yet.
%to have not even been raised yet. 
Hence studies along the lines proposed here, including quantitative comparisons 
between theory and experiments, will prove powerful for future control 
and design strategies, both of unbranched and branched (wrinkle) patterns.
Finally, note that the branching of stripe patterns on the skin of fishes \cite{Kondo:1995.1},
an important example for patterns in living systems, 
is probably also induced by inhomogeneities, similar as discussed here, 
and which are in addition slowly changing with time.

\section*{Acknowledgements}
We thank R. Aichele, A. Fery, W. Pesch (Bayreuth), 
as well as V. Delev (Ufa, Russia) for interesting discussions.

\vspace{6mm}
 
\appendix 
\section{Functionals}
\label{appfunct}

 The dynamical equation Eq.~(\ref{SHanisomod}) for the field $u(x,y,t)$ 
can be derived   from the functional 
\begin{align}\label{funct}
\mathcal{F}[u(x,y,t)]&=\hspace{-1mm}\int \tilde f(u)dx\,dy=\hspace{-1mm}\frac{1}{2}\int\left[-\epsilon u^2+\frac{1}{2}u^4+W\left(\p_x\p_y u\right)^2
 \right. \nonumber\\
&\qquad \left. +c\left(\p_y^2 u \right)^2 + \left(q_0^4(y) u^2+2q_0^2(y)u\Delta u+(\Delta u)^2\right)\right]dx\,dy 
\end{align}
via the functional  derivative
\begin{align}
\p_t u(x,y,t)=-\frac{\delta\mathcal{F}}{\delta u}
 = -\left(\frac{\p \tilde f}{\p u}+\p_x^2\frac{\p \tilde f}{\p(\p_x^2 u)}
+\p_x\p_y\frac{\p \tilde f}{\p(\p_x\p_y u)}+\p_y^2\frac{\p \tilde f}{\p(\p_y^2 u)}\,\right).
\end{align}
The functional in Eq.~(\ref{funct}) can be simplified by assuming %a long system with 
periodic boundary conditions and  by using the expression $\partial_t u$ from the
dynamical equation. 
One gets
\begin{align}
\mathcal{F}[u(x,y,t)]=\frac{1}{2}\int\left[\frac{1}{2}u^4-u(\p_t u)-u^4
-2u(\p_y u)\p_y(q_0^2(y))-u^2\p_y^2\left[q_0^2(y)\right]\right]dx\,dy,
\end{align}
where the $-u^4$-term compensates for the cubic term occurring in $\p_t u$ and similarly
the last two terms.
These last two terms again vanish due to the periodic boundary condition (here in $y$-direction):
\begin{align}
&\int\left[-2u(\p_y u)\p_y(q_0^2(y))-u^2\p_y^2\left[q_0^2(y)\right]\right]dx\,dy
=-\int\p_y\left[u^2\p_y\left(q_0^2(y)\right)\right]dx\,dy=0\,.
\end{align}
We arrive at the simple expression
\begin{align}
\mathcal{F}[u(x,y,t)]=-\frac{1}{2}\int\left[\frac{1}{2}u^4+u(\p_t u)\right]dx\,dy
\end{align}
and if we are -- as in our paper -- interested only in stationary patterns, 
we simply have
\begin{align}
\mathcal{F}[u(x,y,t)]=-\frac{1}{4}\int u^4 dx\,dy\,.
\end{align}
Also the amplitude equation (\ref{Ampli_A}) can be derived from a  functional,
which can be simplified for stationary
solutions in the same manner, only the prefactor changes:
\begin{align}
\mathcal{F}[u(x,y,t)]=-\frac{3}{2}\int |A|^4 dx\,dy\,.
\end{align}

\end{document}